\documentclass{PoS}

\newcommand\be{\begin{eqnarray}}
\newcommand\ee{\end{eqnarray}}
\newcommand{\gsim}{$^>_\sim$}
\newcommand{\mysec}[1]{{\bf #1:}}

\title{Multigrid solver for clover fermions}

\ShortTitle{Multigrid solver for clover fermions}

\author{\speaker{J. C. Osborn}%
\\
 Argonne Leadership Computing Facility,
 Argonne National Laboratory, Argonne, IL 60439, USA
}
\author{R. Babich\\
 Center for Computational Science, Boston University,
 3 Cummington Street, Boston, MA 02215, USA}
\author{J. Brannick\\
 Department of Mathematics, The Pennsylvania State University,
 230 McAllister Building, University Park, PA 16802, USA}
\author{R. C. Brower\\
 Center for Computational Science, Boston University,
 3 Cummington Street, Boston, MA 02215, USA\\
 Department of Physics, Boston University,
 590 Commonwealth Avenue, Boston, MA 02215, USA}
\author{M. A. Clark\\
 Harvard-Smithsonian Center for Astrophysics,
 60 Garden Street, Cambridge, MA 02138, USA}
\author{S. D. Cohen\\
Center for Computational Science,  Boston University,
3 Cummington Street, Boston, MA 02215, USA}
\author{C. Rebbi\\
Center for Computational Science,  Boston University,
3 Cummington Street, Boston, MA 02215, USA\\
Department of Physics, Boston University,
590 Commonwealth Avenue, Boston, MA 02215, USA}

\abstract{
We present an adaptive multigrid Dirac solver developed for Wilson
clover fermions which offers order-of-magnitude reductions in solution
time compared to conventional Krylov solvers.  The solver incorporates
even-odd preconditioning and mixed precision to solve the Dirac
equation to double precision accuracy and shows only a mild increase
in time to solution for decreasing quark mass.  We show actual
time to solution on production lattices in comparison to conventional
Krylov solvers and will also discuss the setup process and its
relative cost to the total solution time.
}

\FullConference{The XXVIII International Symposium on Lattice Field Theory\\
   June 14-19,2010\\
   Villasimius, Sardinia Italy}

\begin{document}

\section{Introduction}
\vspace{-2mm}

Much of the work in lattice QCD goes into the
repeated solution of Dirac equation
\be
\left[ D(U) + m \right] \psi = \eta
\ee
for varying sources $\eta$ and/or gauge fields $U$.
Typically this is done with a Krylov solver such as conjugate gradients (CG)
or BiCGStab.
These methods are known to exhibit critical slowing down
where the time to solution increases rapidly as the quark
mass is decreased.
This is because the condition number of the Dirac matrix tends to diverge
 (for a large enough volume) as the mass decreases ($\kappa \sim 1/m$),
while standard Krylov solvers become inefficient
as the condition number grows (iterations \gsim $\sqrt{\kappa}$).

A lot of work has been done on developing deflation methods that
remove a set of low modes (i.e., eigenvectors with small eigenvalues)
from the solver, leading to improvements in the solver time \cite{defl}.
However all these methods will require deflating a number of modes that
scales linearly with the volume, which then becomes more difficult
as the volume increases.
An alternative is to use the local deflation approach of
 L\"uscher \cite{ldefl} which splits the low vectors into spatial
 blocks which then gives a larger span of modes to deflate, and thus
 does not require a number of vectors that scales with volume.
This splitting of vectors is similar to that done in multigrid methods
\cite{mg} which we describe here.

While multigrid methods have been very successful in other fields,
their application to lattice QCD has been difficult due to the
complexity of the low modes of the QCD Dirac operator.
We have been working to apply the methods of adaptive multigrid
 to QCD starting with plain Wilson quarks in 2d \cite{wmg2d}
 and 4d \cite{wmg4d}.
Here we present the extension to 4d clover improved Wilson quarks
 and show results of a production-ready implementation using
 the USQCD software libraries \cite{usqcd}.

\mysec{Adaptive multigrid}
The main motivation behind multigrid is the observation that in typical
linear systems (from discretized PDEs) the low modes,
which are responsible for the poor convergence of the solver,
are smooth and therefore can be approximated well on a coarser grid
 which reduces the effort required to solve for them.
The problem is then essentially split into two parts with the
 high frequency components being solved on the original (fine) grid
 and the low modes being solved on a coarse grid.
The basic multigrid algorithm consists of alternating between a
 relaxation step (smoother) on the fine grid using traditional iterative
 solvers (typically stationary or Krylov)
 and a solve on a coarse grid.
This procedure can be repeated recursively to solve the coarse grid problem,
reducing to coarser and coarser grids until the coarsest problem is small
enough to be easily solved.

The key components of the algorithm are the choice of smoother and the
operators used for the coarse solve: restriction, $R$, used to project
the error from the fine lattice onto the coarse; prolongation, $P$,
(or interpolation) used to bring the coarse grid correction back up to
the fine level; and the coarse operator itself.  The multigrid
procedure is usually not used on its own and is often used as a
preconditioner for another solver.  For our application, with a
 non-Hermitian Wilson Dirac operator and our choice of a
 non-stationary multigrid cycle, we use generalized conjugate residuals
(GCR) as the outer solver.  Since the basic algorithm for Wilson
fermions has already appeared in \cite{wmg4d}, here we will focus on
the extensions to that algorithm.

\mysec{Red-black preconditioning}
The common approach to solving for the Wilson Dirac operator on the lattice
is to first employ red-black (even-odd) preconditioning which substantially
 reduces the number of iterations needed to solve the system \cite{eo}.
We implement this by splitting the linear system, $D x = b$,
 into even ($e$) and odd ($o$) space-time sites as
\be
\left( \begin{array}{cc}
D_{ee} & D_{eo} \\
D_{oe} & D_{oo} \\
\end{array} \right)
\left( \begin{array}{c}
x_e \\ x_o
\end{array} \right)
=
\left( \begin{array}{c}
b_e \\ b_o
\end{array} \right)
\Rightarrow
\left( \begin{array}{cc}
1 & D_{eo} D_{oo}^{-1} \\
D_{oe} D_{ee}^{-1} & 1 \\
\end{array} \right)
\left( \begin{array}{c}
D_{ee} x_e \\ D_{oo} x_o
\end{array} \right)
=
\left( \begin{array}{c}
b_e \\ b_o
\end{array} \right) ~.
\label{eq:even_odd}
\ee 
The mass term is now absorbed into $D$.  In the preconditioned form,
$D_p y = b$, on the right side of (\ref{eq:even_odd}), we can
solve the reduced system,
$D_r y_e = b_r$, with
\be 
D_r &=& 1 - D_{eo} D_{oo}^{-1} D_{oe} D_{ee}^{-1} \\
b_r &=& b_e - D_{eo} D_{oo}^{-1} b_o ~, 
\ee 
for the even sites, $y_e = D_{ee} x_e$.
The matrices $D_{ee}$ and $D_{oo}$ are diagonal on the space-time lattice
 so we can easily compute $D_r$ and $b_r$ and reconstruct
 $x_e = D^{-1}_{ee} y_e$ and
 $x_o = D^{-1}_{oo} ( b_o - D_{oe} x_e)$.

On the fine level, we use a GCR solver for the reduced system:
 $D_r y_e = b_r$.
However the coarse operator, $\widehat D = R D_p P$, is a
 projection of the full even-odd preconditioned matrix ($D_p$).
When solving the coarse system we again use even-odd preconditioning
 to solve the reduced coarse system.
The interpolation operator ($P$) is formed from the
 low modes of the the full preconditioned operator $D_p$.
One can easily see that the eigenvectors of $D_p$ and $D_r$ are the same on
 the even sites so that the low modes of the two are related.
For the $\gamma_5$-Hermitian Wilson Dirac operator, we constructed the
 restriction operator $R$ using $R = P^\dagger \gamma_5$.
We also split the vectors that formed $P$ into the 2 chiralities in addition
 to the space-time blocks, explicitly preserving $\gamma_5$-Hermiticity 
 on the coarse level and helping to avoid a nearly singular coarse operator.
For the Wilson clover case, $D_p$ is not $\gamma_5$-Hermitian,
 however we still use the same construction.
Initial tests on smaller Wilson lattices found a 30\% improvement using
 the preconditioning, so we have adopted it for clover lattices too.

\mysec{Implementation}
The multigrid solver has been implemented using the US DOE SciDAC
lattice QCD libraries \cite{usqcd} and in particular it is written in
 the C language version of the QCD data parallel layer QDP/C.
This library has been extended to provide multi-lattice support and
 improved arbitrary $N_c$ support which is used to implement the coarse
 level operators (which look similar to a staggered Dirac operator with
$N_c$ equal to the number of vectors used to form $P$ and $R$).

Since the single precision operations are generally faster than double
 precision, we have implemented the multigrid preconditioner in single
 precision while the outer GCR solver is in full double precision.
In all cases tested so far, single precision is sufficiently accurate
to provide a good preconditioner for the Dirac matrix.
For comparison to conventional Krylov solvers below we have also included
 a mixed precision version of those based on iterative refinement.
Other mixed precision methods such as reliable updates \cite{gpu}
may perform better at lighter masses, but we do not expect this to make a
large difference in the final results.

In typical multigrid implementations, one has a choice of how many
cycles of the coarse solve and smoothing to perform at each level
before going back up to the next finer level.  These choices include
the conventional V-cycle, where no extra cycles are performed, or a
W-cycle, where one extra cycle at the coarsest level is performed, or
some more sophisticated pattern.  Here instead of choosing the pattern
up front, we implement the recursive solver in a truly adaptive
fashion where the coarse grid solver is again a GCR solver
preconditioned with another multigrid cycle, and so on until the
coarsest lattice, which is solved with an unpreconditioned GCR.  At
each level we only specify the residual tolerance for each GCR solver
which then lets the solver at each level determine how many cycles are
needed to reach the requested tolerance.  We ran the multigrid solver
with many different sets of parameters and found that a tolerance of
around 0.1 is usually best.  This means that the coarse solves do not
need to be done very accurately.  In all results below the times
reported were for the best set of parameters found for that particular
case.

\begin{figure}[t]
  \vspace{-5mm}
  \begin{minipage}[t]{.5\textwidth}
    \begin{center}
      \includegraphics[width=1\columnwidth,clip]{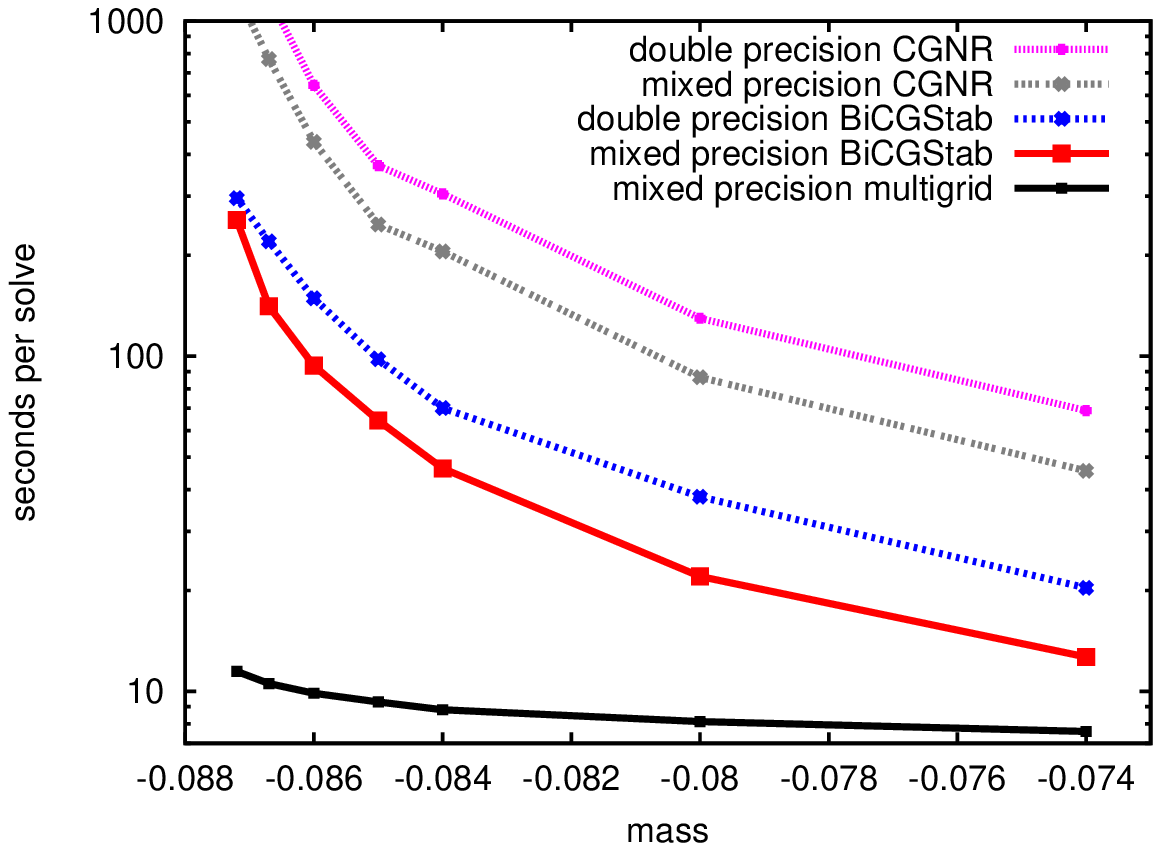}
    \end{center}
  \end{minipage}
  \vspace{-5mm}
  \begin{minipage}[t]{.5\textwidth}
    \begin{center}
      \includegraphics[width=1\columnwidth,clip]{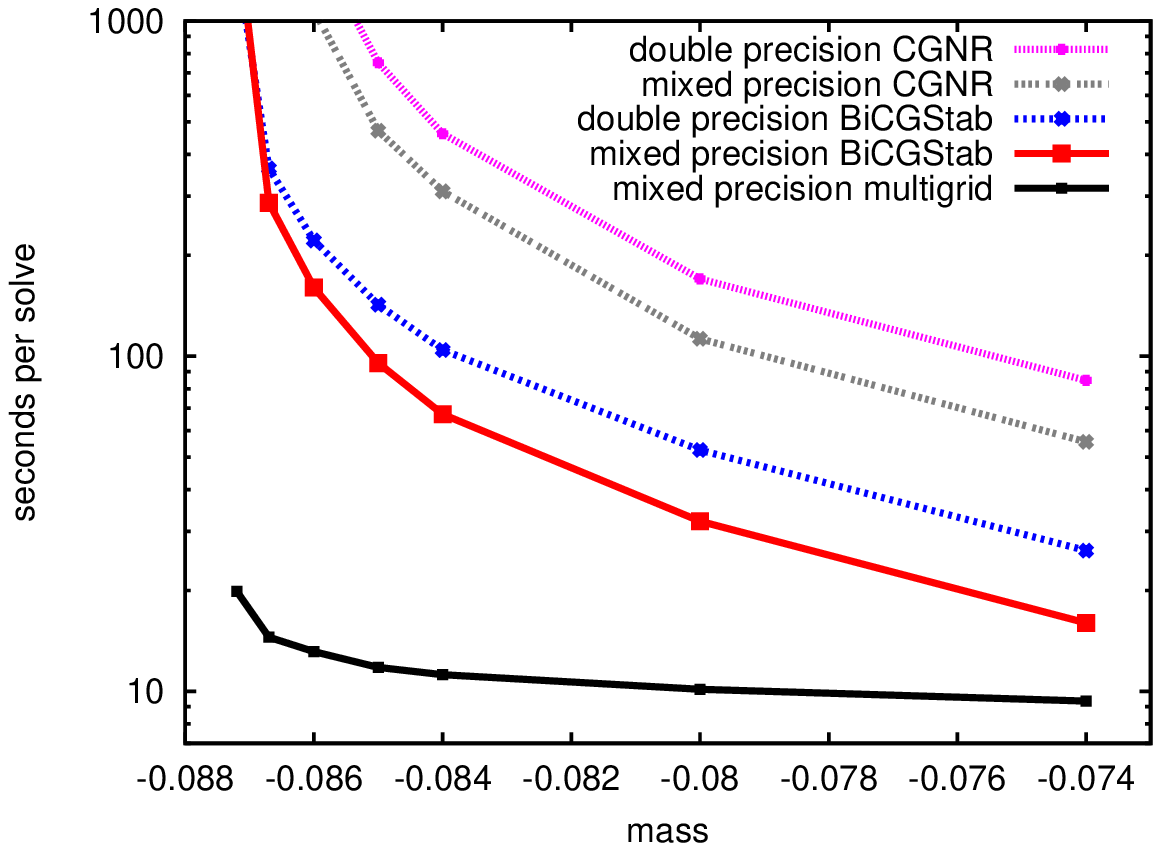}
    \end{center}
  \end{minipage}
  \vspace{-4mm}
  \caption{Time to solution (in seconds) for a single solve versus
           quark mass for various Krylov solvers and multigrid.
           The left plot is for the $24^3\times 128$ lattice and
           the right is for $32^3\times 256$.}
  \vspace{-4mm}
  \label{figtvm}
\end{figure}

\vspace{-5mm}
\section{Results}
\vspace{-3mm}

We have tested the multigrid algorithm on the 
Hadron Spectrum Collaboration anisotropic clover lattices \cite{hsc}.
We have run on lattices of size $24^3\times 128$ and $32^3\times 256$
with spatial and temporal lattice spacings of
$a_s \approx 0.12$ fm and $a_t \approx 0.035$ fm.
The light dynamical mass ($m_l = -0.086$) corresponds to
$m_\pi \approx 220$ MeV.
For reference the strange quark on these lattices was determined to be
$m_s \approx -0.0743$, while the physical light quark mass would be around
$m_l \approx -0.0867$.
The sizes of the coarse lattices and number of low vectors used in the
restriction and prolongation operators are given in table \ref{tabmg}.
All results were obtained on 256 or 1024 cores of a Blue Gene/P.

\begin{table}[b]
\begin{tabular}{|c|c|c|c|c|c|}
\hline
lattice & cores &
 $1^{st}$ coarse lattice & \# vectors in $1^{st}$ $P$ &
 $2^{nd}$ coarse lattice & \# vectors in $2^{nd}$ $P$ \\
\hline
$24^3\times128$ & 256 &
 $8^3\times16$ & 24 &
 $4^3\times4$ & 32 \\
\hline
$32^3\times256$ & 1024 &
 $16\times8\times8\times32$ & 24 &
 $4^3\times16$ & 32 \\
\hline
\end{tabular}
\caption{multigrid parameters}
\label{tabmg}
\end{table}

Figure \ref{figtvm} shows the time to solution for a single solve
 as a function of the quark mass for various solvers for the
 two lattice volumes.
In all cases the conjugate gradients on the normal residual (CGNR)
 performed worse than BiCGStab and the mixed precision solver performed
 better than pure double precision.
The multigrid solver performs much better than the others and shows
 a much smaller increase in time as the quark mass is decreased.
For the larger volume the speedup factor of multigrid compared to
 mixed precision BiCGStab is about $1.7\times$ at the heaviest mass,
 $12.2\times$ at the dynamical light mass
 and $19.8\times$ at the physical light mass.
On the larger volume we see a sharper increase in time at the smallest
 masses compared to the smaller volume.
We expect that this could be improved with
 additional work in the setup and/or adding a fourth level.

\begin{figure}[t]
 \vspace{-5mm}
 \begin{minipage}[t]{0.5\textwidth}
  \begin{center}
   \includegraphics[width=1.0\columnwidth,clip]{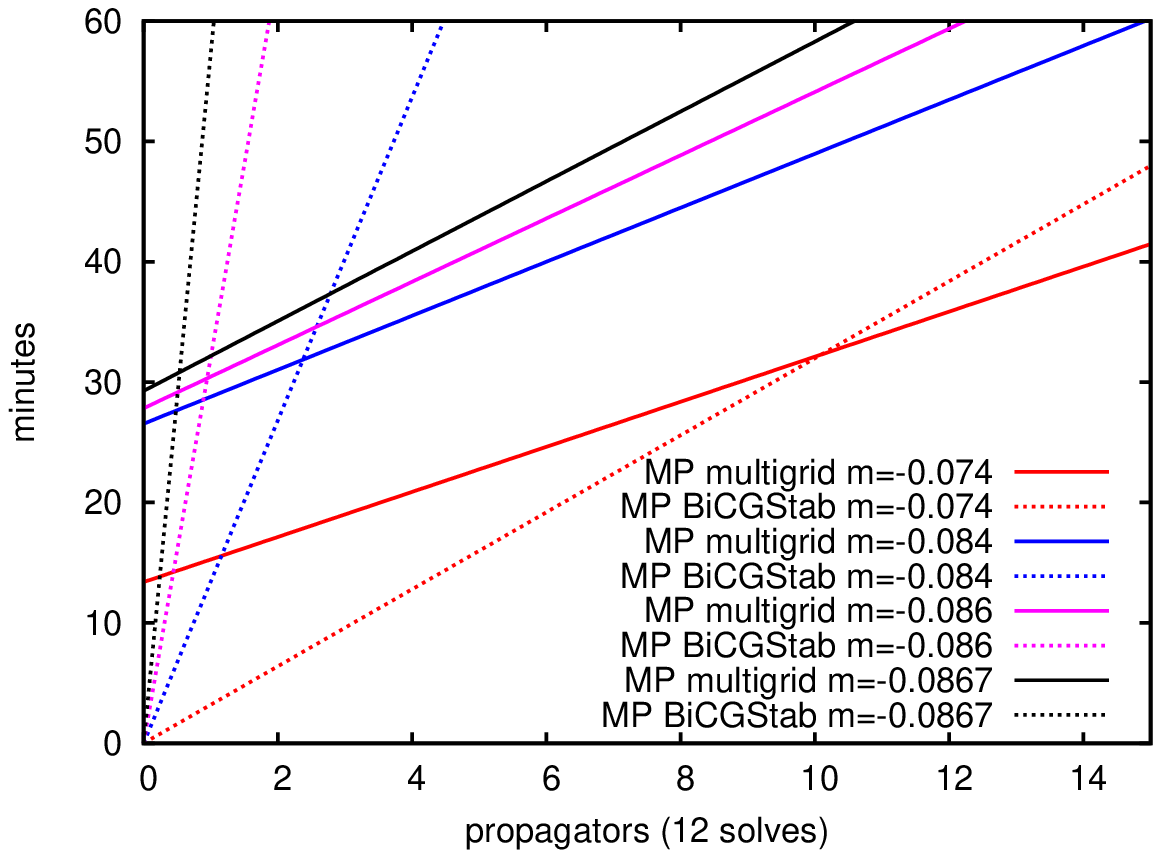}
   \vspace{-8mm}
   \caption{Total time to solution including setup versus number of right
            hand sides for mixed precision (MP) BiCGStab and multigrid
            at different quark masses.
          }
   \label{figttvsolve}
  \end{center}
 \end{minipage}
 \vspace{-5mm}
 ~~~
 \begin{minipage}[t]{0.5\textwidth}
  \begin{center}
   \includegraphics[width=1.0\columnwidth,clip]{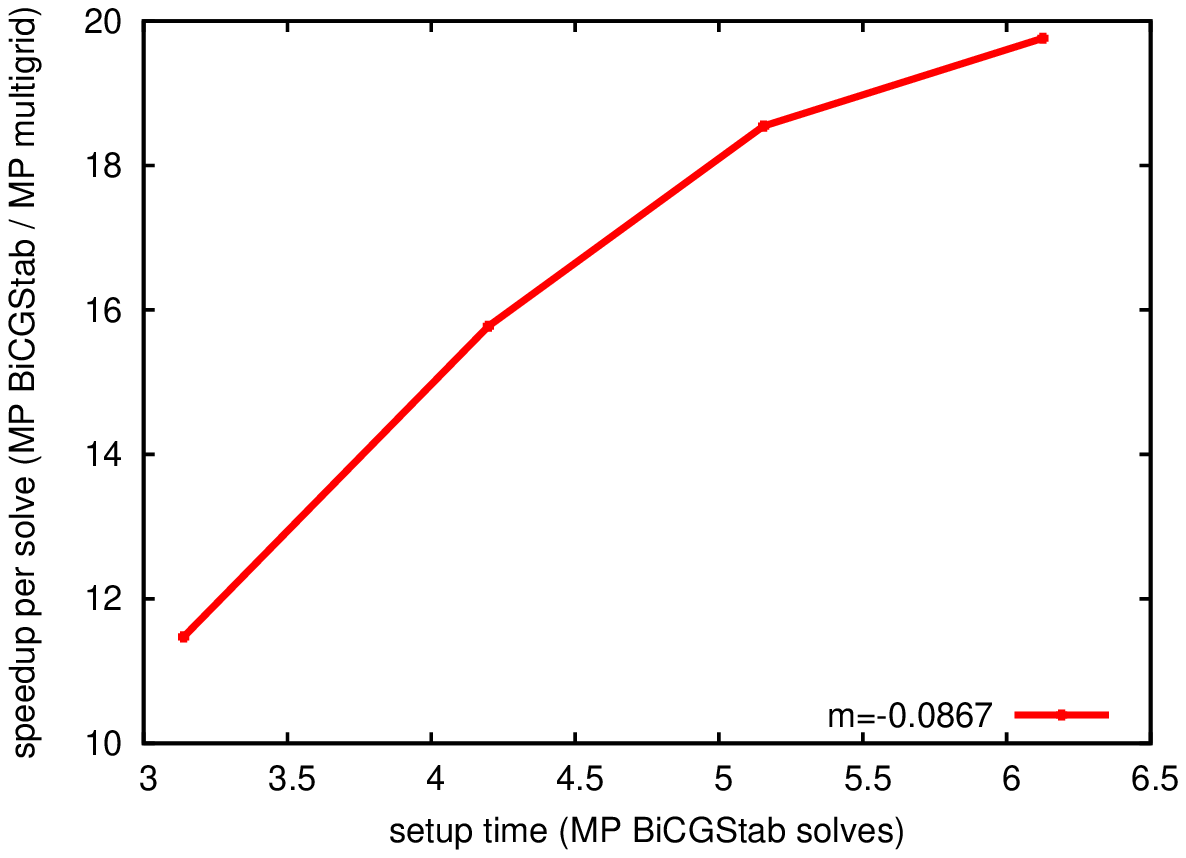}
   \vspace{-8mm}
   \caption{Speedup of multigrid solver relative to BiCGStab versus
            setup time at the physical quark mass.}
   \label{figtvs}
  \end{center}
 \end{minipage}
\end{figure}

In figure \ref{figttvsolve} we show the total time required to solve
different number of right hand sides, including the setup cost,
on the larger volume.  For
small numbers of solves, BiCGStab is faster due to the setup time
needed for multigrid.  As the number of solves increases multigrid
becomes faster due to the improved time per solution.  The break even
point becomes smaller as the quark mass is decreased.
At close to strange quark mass the crossing is at around 10 full
 propagators (of 12 solves each).
At the dynamical mass the crossing is at 1 full propagator and
at the physical mass it is about half a propagator (6 solves).
Of course it is possible to save the vectors and even the coarse
matrix to load back in for later analysis, so for analysis projects on
saved configurations, the setup cost should not be an issue.  Only for
configuration generation is the setup cost an issue.
Since the main focus for the implementation is currently for analysis,
the setup code has not been fully tuned and there is still room for
improvement both algorithmically and in code optimization.

One still has some freedom to choose how much time to spend in the
setup procedure which then affects the quality of the resulting
solver.
In figure \ref{figtvs} we plot the speedup for a single application of
the multigrid solver relative to BiCGStab versus the time
spent in the setup (in units of the time for a single BiCGStab solve).
These results were obtained on the larger lattice at the physical
quark mass.  If we spend about 6 BiCGStab solves worth of work in the
setup we get a solver that is about $20\times$ faster than BiCGStab,
which is what was used in the previous figures.
If we lower the setup cost to about 3 BiCGStab solves, then the solver
speedup reduces to around $11\times$.

\begin{figure}[t]
 \vspace{-5mm}
 \begin{minipage}[t]{0.5\textwidth}
  \begin{center}
   \includegraphics[width=1.0\columnwidth,clip]{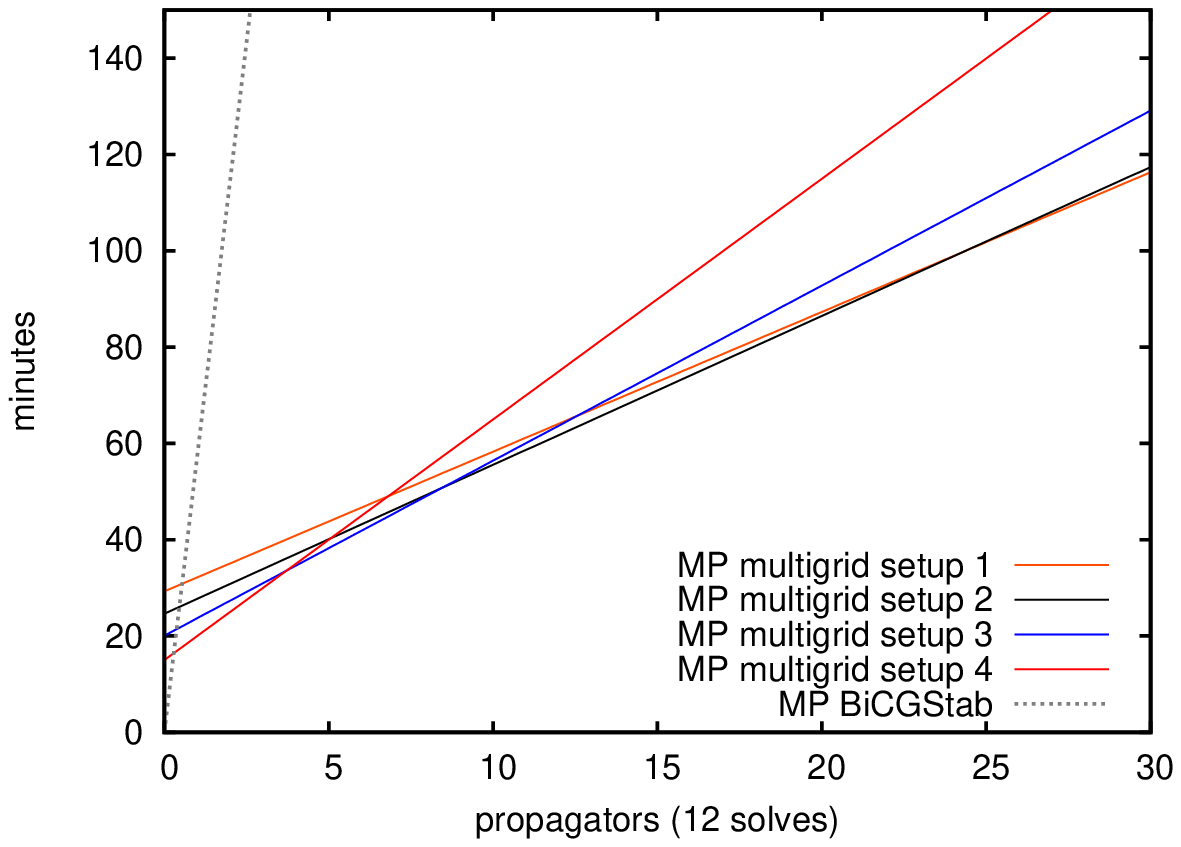}
   \vspace{-8mm}
   \caption{Total time versus number of solves for different amounts
            of setup work on the $32^3\times256$ lattice at the physical mass.}
   \label{figttvsetup}
  \end{center}
 \end{minipage}
 \vspace{-5mm}
 ~~~
 \begin{minipage}[t]{0.5\textwidth}
  \begin{center}
   \includegraphics[width=1.0\columnwidth,clip]{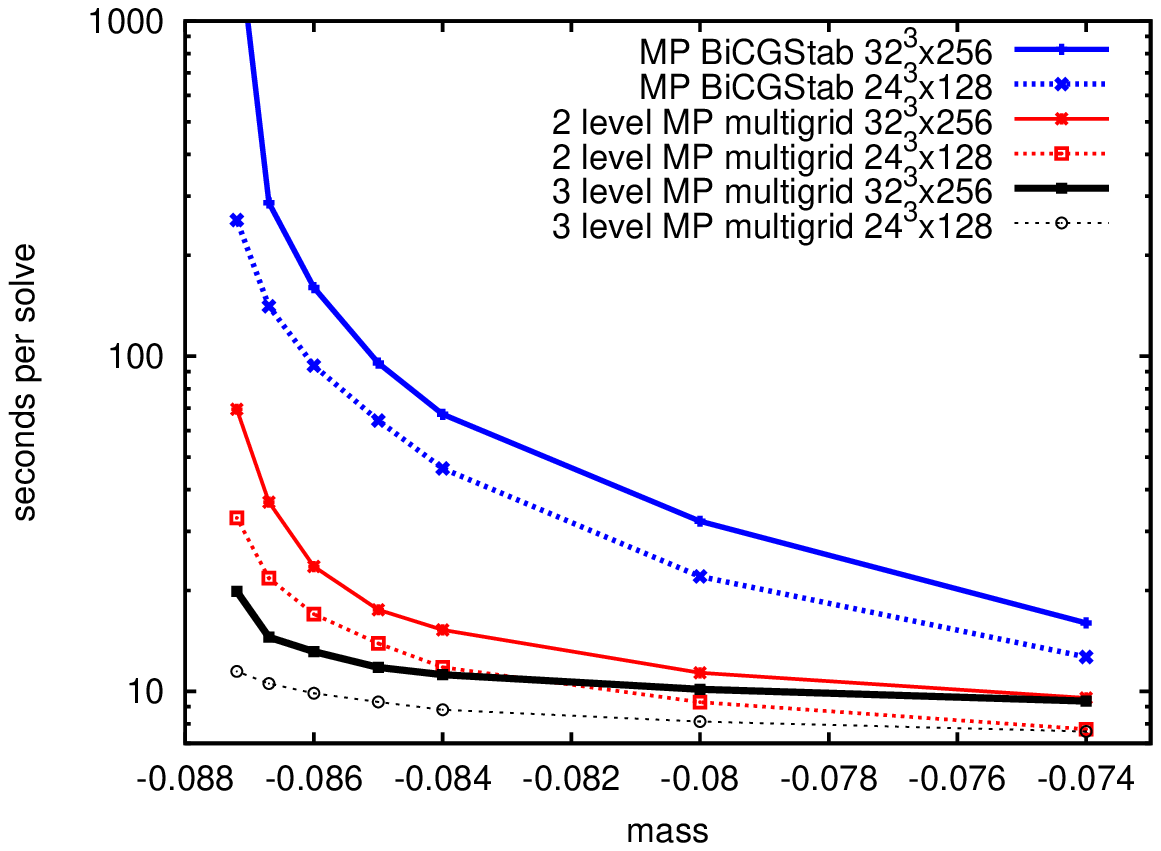}
   \vspace{-8mm}
   \caption{Comparison of 2 and 3 level multigrid algorithms on
            both lattice sizes.
          }
   \label{fig2v3}
  \end{center}
 \end{minipage}
\end{figure}

We can see how this freedom can be used to optimize the total time in
 figure \ref{figttvsetup}.
Here we show the total solution time including setup versus number of solves
 for the four different setups shown in the previous figure.
These runs were again done on the larger lattice at the physical mass.
Here we see that the smallest setup time gives the best total performance up to
 about 4 propagators at which point the second smallest setup becomes best.
The third setup takes over at around 8 propagators and the last at around 25.
Thus if the setup is not being saved for reuse at a later time, one can 
optimize the setup for the particular work being done.

In figure \ref{fig2v3} we compare the performance of the 2-level and 3-level
multigrid algorithms for both lattice sizes.
For heavier masses the difference between 2 and 3 levels is small
while both are still better than BiCGStab.  For lighter masses the 3
level algorithm is clearly better and is about $2.5\times$ better at the
physical quark mass.
As noted earlier the increase in time seen for the 3-level algorithm
at the lightest quark masses suggests that 
improving the coarse level with additional work in the setup and/or 
adding a fourth multigrid level may be beneficial here.

\begin{figure}[t]
 \vspace{-5mm}
 \begin{minipage}[t]{0.5\textwidth}
  \begin{center}
   \includegraphics[width=1.0\columnwidth,clip]{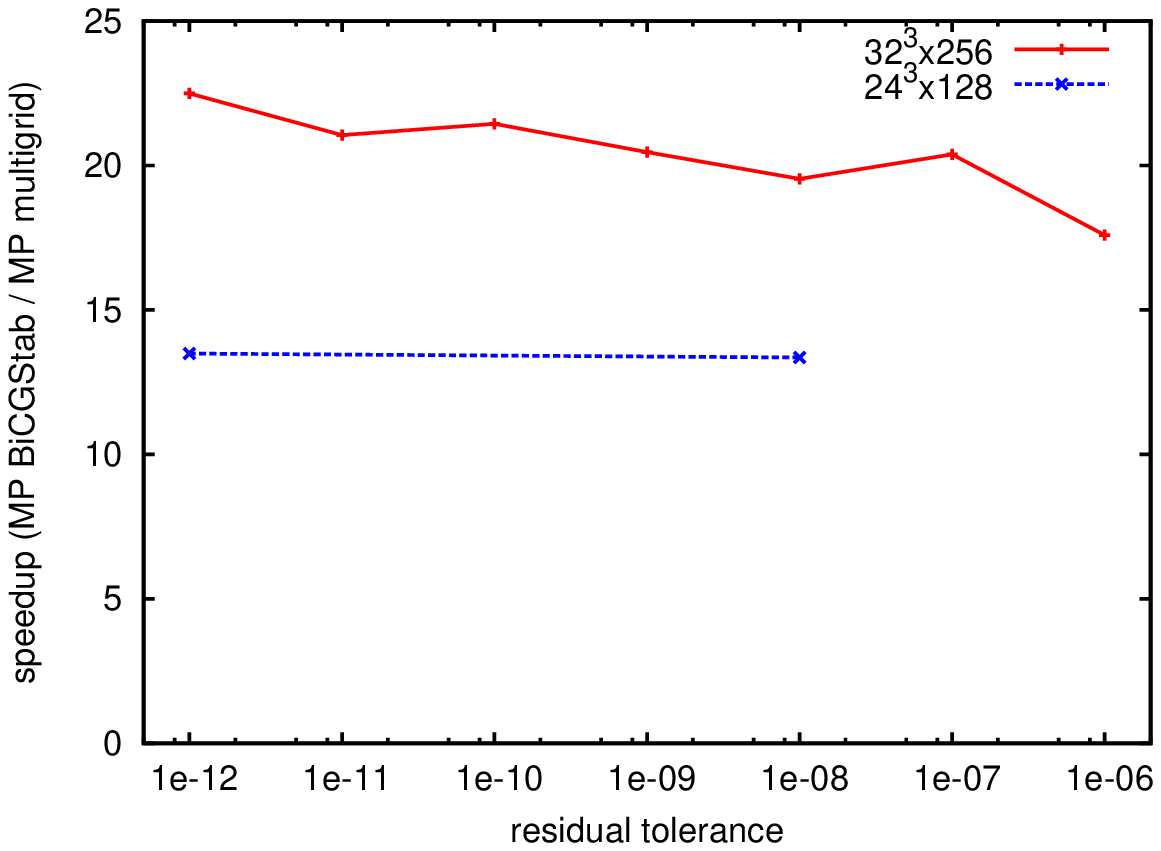}
   \vspace{-8mm}
   \caption{Speedup of multigrid over BiCGStab versus requested
            residual tolerance.  On the larger lattice
            we see the speedup increase as the tolerance
            decreases.}
   \label{figmgr}
  \end{center}
 \end{minipage}
 \vspace{-5mm}
 ~~~
 \begin{minipage}[t]{0.5\textwidth}
  \begin{center}
   \includegraphics[width=1.0\columnwidth,clip]{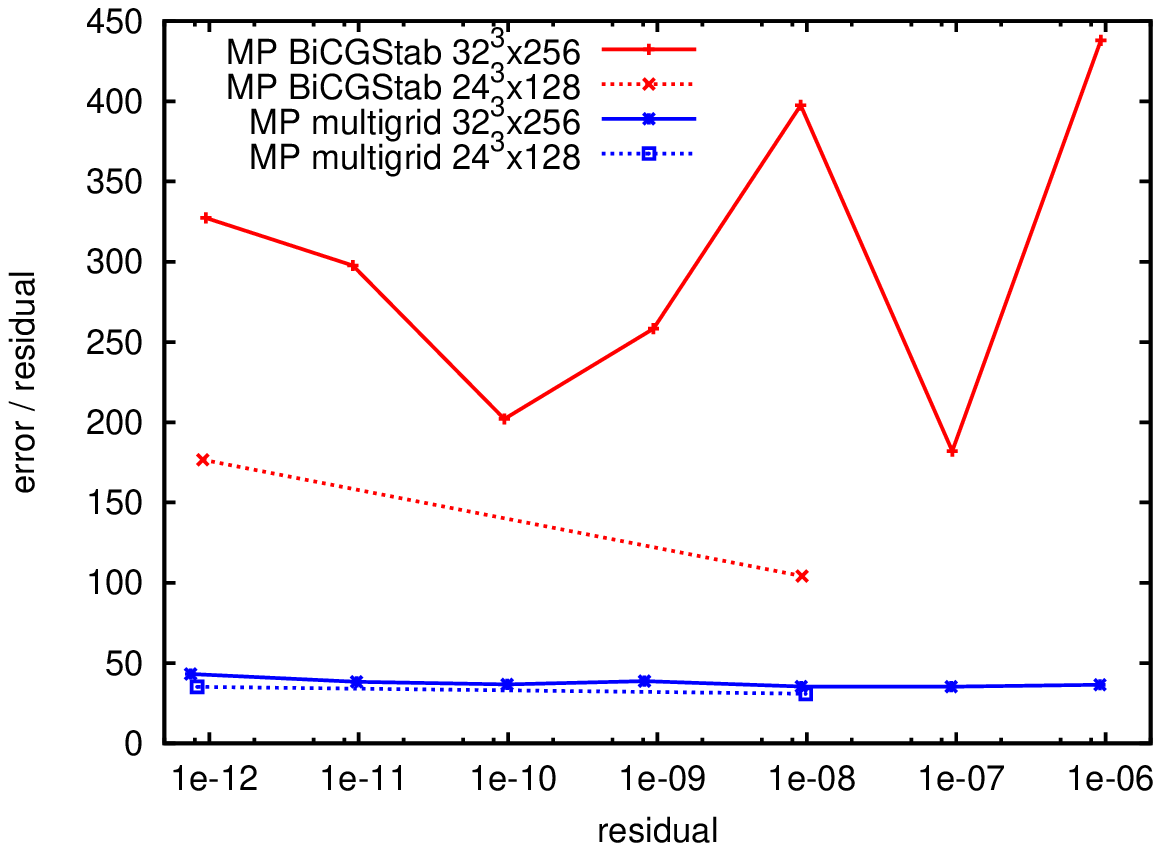}
   \vspace{-8mm}
   \caption{Ratio of magnitude of error to magnitude of residual
            versus magnitude of residual for BiCGStab and multigrid.
          }
   \label{figerr}
  \end{center}
 \end{minipage}
\end{figure}

In figure \ref{figmgr} we show how the relative speedup of multigrid over
BiCGStab varies with the requested residual tolerance.
These results are obtained at the physical mass.
For the smaller volume the speedup appears to stay constant as the tolerance
changes, however for the larger volume, the speedup tends to increase
as the tolerance is lowered.
Thus the multigrid algorithm is even more effective for smaller
tolerances.

Although the residual is typically used as the measure of convergence
due to it being readily available, what usually matters for
observables is the actual error defined by
$e = D^{-1} b - x$.
It is related to the residual $r$ by
$r = b - D x = D e$.
Since the residual is the error multiplied by $D$, it is not as
 sensitive to low modes.
In figure \ref{figerr} we plot the ratio of the magnitude of the
error to the magnitude of the residual versus the magnitude of the
residual for BiCGStab and multigrid at the physical mass.
In order to know the exact solution, we first take a point source ($p$),
 then solve against that to get an approximate solution
 $x_0 \approx D^{-1} p$.  We set the right hand side to be
$b = D x_0$ so that $b$ is approximately a point source and its exact
solution is known.
The error is very stable for multigrid and stays at about
$40-50\times$ the residual for both lattice volumes.  The BiCGStab
error fluctuates wildly at about $5-10\times$ larger than the
multigrid error and appears to grow with volume.

\vspace{-3mm}
\section{Setup procedure}
\vspace{-2mm}

Currently the setup procedure consists of a sequence of repeated relaxations
 (inverse iteration) on a random vector, $v$, while monitoring the Rayleigh
quotient
$v^\dagger D^\dagger D v/v^\dagger v$
to determine when the vector has converged well enough onto the 
low modes of the system.
During this process we also keep the current vector globally orthogonal to the
previously converged vectors.
The main motivation for implementing this setup procedure is its simplicity
 since one doesn't need to construct coarse operator until all the vectors
 are found.
It is also relatively easy to vary the number of iterations and the convergence
criteria to tune the cost of the setup and consequently the efficiency of the
resulting solver.
However the main drawback of this procedure is that the resulting vectors may
be locally redundant within the blocks.
More sophisticated setup procedures have been developed to avoid
 this problem.
One such setup procedure is used in the adaptive smooth aggregation
 ($\alpha$SA) method \cite{asa}.
Here one constructs a multigrid cycle out of the currently available vectors
and uses that solver to relax on random vectors which will then give a new
 vector which is rich in the modes that the current solver is bad at resolving.
This requires construction of coarse operator several times during the setup
process which adds to the complexity and possibly also the time of the setup.
A hybrid approach combining these procedures was developed for the Wilson case
 with promising results \cite{wmg4d}.
We plan to implement this for the clover case in the future.

\vspace{-3mm}
\section{Summary and plans}
\vspace{-2mm}

We have developed an efficient implementation of a Wilson clover
multigrid solver, currently being used in production for the
calculation of disconnected diagrams.
For light quarks we see a $10-20\times$ reduction in time to solution.
We also note that the error is very stable and relatively small compared
to Krylov solvers and that the speedup and relative error improves for
larger lattices.  We are now in the process of testing it on larger
lattices and working on optimizing the code so it can be extended to
even coarser lattices.  Currently the solver is a great improvement
for medium to large analysis projects where the setup cost can be
amortized over many solves.  We are working to improve the setup
process to provide the same or better quality solver and lower cost so
that it can be readily used in smaller projects or in configuration
generation. Finally we are working on multigrid for domain wall and
staggered quarks, as well as porting these solvers to GPUs.

\mysec{Acknowledgments}
This work was supported by the US NSF and DOE.
Runs were performed at the Argonne Leadership Computing Facility
which is supported under DE-AC02-06CH11357.

\end{document}